\documentclass{IEEEtran}
\usepackage{url,epsfig,textcomp,epstopdf,algorithmic,algorithm,color}
\usepackage{amsbsy,amsmath, amssymb,comment,amsthm,dsfont}
\usepackage{pdfpages}
\usepackage{mathtools}
\newtheorem{definition}{Definition}\newtheorem{theorem}{Theorem} \newtheorem{corollary}{Corollary}
\newcommand{\bth}{\begin{theorem}}\newcommand{\ethe}{\end{theorem}} \newcommand{\bpr}{\begin{proof}}\newcommand{\epr}{\end{proof}} \newcommand{\ble}{\begin{lemma}}\newcommand{\ele}{\end{lemma}} \newcommand{\bco}{\begin{corollary}}\newcommand{\eco}{\end{corollary}}
\newcommand{\bde}{\begin{definition}}\newcommand{\ede}{\end{definition}}
\newcommand{\m}{\mathcal}
\title{Obfuscation using Encryption}

\author{\IEEEauthorblockN{Johannes Schneider\IEEEauthorrefmark{1} and Thomas Locher\IEEEauthorrefmark{2}}\\
	    \IEEEauthorblockA{\IEEEauthorrefmark{1} University of Liechtenstein, Liechtenstein
	    	\\\{firstname.lastname\}@uni.li}\\
	    \IEEEauthorblockA{\IEEEauthorrefmark{2} ABB Corporate Research, Baden-Daettwil, Switzerland
	    	\\\{firstname.lastname\}@ch.abb.com}
	}
	
\begin{document}

\maketitle

\begin{abstract}
Protecting source code against reverse engineering and theft is an important problem. The goal is to carry out computations using confidential algorithms on an untrusted party while ensuring confidentiality of algorithms. This problem has been addressed for Boolean circuits known as `circuit privacy'. Circuits corresponding to real-world programs are impractical. Well-known obfuscation techniques are highly practicable, but provide only limited security, e.g., no piracy protection. In this work, we modify source code yielding programs with adjustable performance and security guarantees ranging from indistinguishability obfuscators to (non-secure) ordinary obfuscation. 
The idea is to artificially generate `misleading' statements. Their results are combined with the outcome of a confidential statement using encrypted \emph{selector variables}. Thus, an attacker must `guess' the encrypted selector variables to disguise the confidential source code. We evaluated our method using more than ten programmers as well as pattern mining across open source code repositories to gain insights of (micro-)coding patterns that are relevant for generating misleading statements. The evaluation reveals that our approach is effective in that it successfully preserves source code confidentiality.

\end{abstract}

\section{Introduction}\label{sec:intro}
Intellectual property, e.g., in the form of algorithms, is costly to develop and it is always at risk of being stolen. In particular, in an untrusted cloud environment, algorithms holding valuable expert knowledge are susceptible to theft: If the cloud infrastructure is compromised, an attacker can steal (compiled) program code. Insider attacks are also a real threat. A typical approach to (partially) protect intellectual property is to obscure it in such a way that it becomes difficult to figure out its purpose and functionality. However, such obfuscation does not solve the problem satisfactorily as the functioning of the algorithm can still be observed and analyzed, even to the point where the whole algorithm and its parameters are reconstructed. Thus, an attacker with access to the cloud can still steal an algorithm and execute it without modification. In turn, Boolean circuits can be cryptographically protected, e.g., using fully homomorphic encryption. An attacker obtaining a `protected' circuit cannot evaluate it. 
However, computations on Boolean circuits alone are considered highly impractical due to the large size of circuits that are needed for complex functionality. We tackle the problem at a higher level by encrypting statements in source code instead of individual gates. Our primary focus is on encryption of source code as written by the programmer (or intermediate code). Our obfuscation technique yields analogous guarantees as for Boolean circuits, i.e., candidate indistinguishability obfuscators, using higher level source code primitives. Moreover, we discuss how to obtain less secure obfuscations that run much faster. Obfuscation is done by a novel technique: we transform source code into ``encrypted'' source code by adding a multitude of additional misleading statements (and variables) combined with selector variables. Selector variables are encrypted binary variables that ``choose'' the right result among misleading and confidential statements. 
Since the selector variables are encrypted, it is generally non-trivial to determine which statements merely serve the purpose of misleading an attacker and which statements actually contribute to the computation of the result. In particular, current de-obfuscation methods and tools are unable to do so. The difficulty of de-obfuscation or extracting the confidential source code depends on the background knowledge of the attacker about the program to de-obfuscate as well as on the choice of the misleading statements. 
We also provide an analysis of several open source code programs to determine patterns that might have to be observed when creating obfuscated source code. Finally, we give a brief assessment of our method using a small-scale experiment using actual programmers whose task was to break a simple encrypted piece of code. Programmers were very far from breaking our obfuscation scheme despite the fact that we used a rather limited degree of source code encryption. Thus, the results provide some evidence that the proposed method is indeed effective.

\section{Model and Problem}
A client wishes to execute confidential source code in an untrusted environment. The data might or might not be confidential, but we assume that the source code computes on encrypted data. Typically, the clients encrypt code (and data) and send both to an untrusted server. The server performs possibly multiple computations using the encrypted code and various input data. It returns encrypted results to clients, where the result are decrypted. 

Intuitively, the mechanism is secure if the adversary having full access to the server (including CPU registers, memory, hard drives and the obfuscated code) does not learn anything about the source code or about the data. However, in practice, it might be sufficient if only certain parts or aspects of a piece of software remain confidential, eg. the call graph or the architecture. We focus on the case where an attacker should learn as little as possible how the computation of the output works.

We assume that all encryptions are perfectly secure, i.e., an attacker cannot gain any knowledge from encrypted data. 

Obfuscation maps a confidential program $P_C$ from a program class $\mathbb{C}_{C}$ to another program from a program class $\mathbb{C}_{O}$, ie. we can express an obfuscator $O$ as  $O: \mathbb{C}_{C} \rightarrow \mathbb{C}_{O}$. As we shall discuss later, the relationship is non-injective, ie. every program $P_O \in \mathbb{C}_{O}$ maps to some set $\mathbb{C}_{C}$. Whereas operations in the given program $P_{C}$ operate on plaintexts, obfuscated programs in $\mathbb{C}_{O}$ perform computations on encrypted data. For example, a program $a\cdot b$ with plaintext variables $a,b$ might correspond to $ENC(a)\cdot ENC(b)$ on encrypted values, ie. $ENC(a), ENC(b)$ denotes ciphertexts and the `$\cdot$' operator performs multiplication of plaintexts using encrypted values only, such that $DEC(ENC(a)\cdot ENC(b)) = a\cdot b$, where $DEC$ denotes decryption. For this example, knowing $ENC(a)\cdot ENC(b)$ suffices to discover the program on plaintexts, ie. $a\cdot b$, since the performed operations on plaintext can generally be inferred by the attacker having the obfuscated program with operations on ciphertexts. For example, for the Goldwasser-Micali crytposystem multiplication of ciphertexts corresponds to an XOR of plaintexts. We assume that the attacker knows which cryptoysystem is employed. For more general computations we need either fully homomorphic encryption or secure multi-party computation.  Generally, merely encrypting data used for computation does not provide any security of algorithms.\\ 

The adversary might know the obfuscation algorithm and program classes $\mathbb{C}_{O}$ and $\mathbb{C}_{C}$. The task of the attacker is to assign to each program $P$ from the program class $\mathbb{C}_{C}$ a probability $p(P)$ stating his belief that the program $P$ is the (or a) confidential program $P_C$. Thus, the goal of the attacker is to choose $P$ to maximize $p(P)$ without knowing the distribution $p$. However, an attacker might have some knowledge about the distribution. In particular, he might know coding patterns, eg. covering a few lines of code, that are more likely to occur than other lines of code. Generally, finding a program $P'$ that differs only slightly from the confidential program $P_C$ might also be satisfactory. An attacker is said to successfully break our scheme, if it can do so using any computation taking polynomial time in the size of the largest program in $\mathbb{C}_{O}$.

\subsection{Indistinguishability Obfuscation}
Indistinguishability obfuscation (see Definition 1 in \cite{gar13}) essentially says that the obfuscations of two circuits look `equivalent' to an attacker. This definition focusing on circuits can be extended to more general programs being composed of more complex commands than Boolean gates. By contrast, we also  use additional \emph{misleading} input data for a program that has no impact on the result, eg. a function $f(x)=x^2$ can be extended to $f(x,y)=x^2+0\cdot y$. Therefore, we assume that any program can be called with any superset of the required input to compute the result.  A program class $\mathbb{C}_{C}$ is an arbitrary set of programs. We discuss program classes and their relation to our obfuscation technique later (see Section \ref{sec:procla}). More formally, we define similarly to Definition 1 in \cite{gar13}: 
\begin{definition}[Indistinguishability Obfuscator (iO)  ]\label{def:iO}
A uniform Probabilistic Polynomial-Time  (PPT)  machine iO: $\mathbb{C}_{C} \rightarrow \mathbb{C}_{O}$ is called an indistinguishability obfuscator for program class $\mathbb{C}_{C}$ if the following is satisfied:\\
i) For all $P \in \mathbb{C}_{C}$ and inputs $x$, we have
$$Pr[DEC(P'(ENC(x)))= P(x): P' \leftarrow iO(P)] =1 $$\\
ii) For any PPT distinguisher D, there exists a negligible value $\epsilon$ such that for all pairs of programs $P,P' \in \mathbb{C}_{C}$ if $P(x)=P'(x)$ for all inputs $x$, then
$$|Pr[D(iO(P))=1] - Pr[D(iO(P'))=1]| \leq \epsilon$$
\end{definition}
 Condition (i) essentially demands that the obfuscated program and the non-obfuscated program produce the same outcome. Condition (ii) says that knowing an obfuscated program is of no use to an attacker.
This definition by itself is not sufficient to ensure confidentiality without using proper program classes. In other words, an obfuscator satisfying the above definition does not necessarily protect an algorithm well. In particular, the confidential program is not protected sufficiently if the program class is not meaningful in that many programs in the program class can easily be ruled out as the confidential program (see Section~\ref{sec:procla}). Furthermore, Condition (ii) in the above definition (and also in \cite{gar13}) focuses  only on avoiding the leakage of programs that behave identically with respect to outputs, ie. $P(x) = P'(x)$. In practice, any program with very similar input-output behavior to the confidential program might also be deemed confidential.

\subsection{Program Definition}
We keep to a simple yet sufficient definition. Assume that a program $P$ of length $n$ is a sequence of statements $s_0,s_1,s_2,...,s_{n-1}$. A statement $s_i$ is an assignment of a (simple) expression to a variable, ie. a statement $s$ could be $r:=a\cdot b$ with expression $a\cdot b$ and result variable $r$. A simple expression consists of a single operation and two (input) variables. We generally write it in the form $Op(Input1,Input2)$, eg. $MUL(a,b)$ denotes the multiplication $a \cdot b$. The program returns the result, ie. variable, of the last statement. Later (see Definition \ref{def:comb}), we introduce in more detail \emph{combining statements}, which aggregate the results of several misleading statements and one confidential statement. Note that this simple definition of a statement also covers control-flow statements for code operating on encrypted data (only), since conditional statements must be transformed to hide the outcome of the branching conditions. For example, loop unrolling is done using an upper bound on the possible number of iterations. \\

For illustration, consider a program to square a number $a$. Let program $P(a)$ consist of a single statement $b:=MUL(a,a)$ with input value $a \in \{0,1,...10\}$. Assume the set of possible operations in the used programming language is given by $S_O:=\{MUL,ADD,DIV\}$ and there is just one variable $a$. A simple expression consists of an operation and its input variables, eg., $MUL(a,a)$. If we define the set of programs to be all programs with one simple expression and one input variable, we could define the program class as $\m{P}=\{(b:=MUL(a,a)),(b:=ADD(a,a)),(b:=DIV(a,a))\}$. If we use two statements, where the first defines some variable, we get $\m{P}=\{(b:=MUL(a,a);c:=MUL(a,b)),(b:=MUL(a,a);c:=ADD(a,b)),...\}$. Out of these programs three actually perform the desired functionality of squaring a number, namely $(b:=MUL(a,a);c:=MUL(a,a)) , (b:=ADD(a,a);c:=MUL(a,a)), (b:=DIV(a,a);c:=MUL(a,a))$. The number of simple expressions given $|V|$ variables and $|S_O|$ operations taking $t$ input variables can be 
bounded by $|S_O|\cdot |V|^t$. 
Any of the prior programs that squares a number might serve as an obfuscated program in the traditional sense (i.e., without using encryption).  But standard dead code elimination would already remove the added misleading statement for obfuscation. Next, we introduce a more sophisticated technique that makes use of encrypted selector variables.

\section{Source Code Obfuscation through Encryption}
We distinguish two levels of source code encryption differing in the unit of abstraction they seek to conceal: Program level encryption considers programs as a whole.  A finer granularity is given by hiding individual statements.  


\subsection{Program Level} \label{sec:pro}
The idea is to conceal the confidential program is to run (all) possible programs (not just those yielding the desired output), ie. all programs from class $\mathbb{C}_{C}$. Then we select the confidential result out of all results using encrypted binary selector variables. Since all programs are run and only the result is selected, an attacker obtains no information which of the executed programs yields the result. Thus, the class $\mathbb{C}_{O}$ consists of just one program executing all programs in $\mathbb{C}_{C}$ using encrypted inputs. It is important that the order of computation of the programs is randomized, ie. we randomly permute all programs and then evaluate them.\\
Let $ENC(r_i)$ be the encrypted result of program $P_i(x) \in \mathbb{C}_{O}$ for an input $x$. Define $b_i \in \{0,1\}$ to be a selector bit for program $i$, ie. $b_i=1$ if $i=i^*$ and 0 otherwise. Let $P_{i*}=P_C$ be the confidential program, e.g., with the functionality to square a number. We say $P_{i^*}(x)=r^*$ yields the desired result $r^*$. The final encrypted result $r^*$ can be obtained by $ENC(r^*):= \sum_{i} ENC(b_i)\cdot ENC(r_i)$, where `$\cdot$' is a multiplication on encrypted data and the summation is also performed on encrypted values. The client can decrypt $ENC(r^*)$ to get $r^*$. For example, say  $P_0=P_C:=\{MUL(a,a)\}$ and $P_1:=\{\}ADD(a,a)\}$. We set $b_0=1$ and $b_1=0$. Therefore, $ENC(r^*):=ENC(MUL(a,a))\cdot ENC(b_0) + ENC(ADD(a,a))\cdot ENC(b_1)=ENC(MUL(a,a))$

Since the bits $b_i$ as well as the results are encrypted, an attacker does not know which of the results is chosen. It is immediate that such an obfuscator satisfies Definition~\ref{def:iO} given that all secure computation mechanisms are secure, ie. we fulfill condition (i) due to the definition of selector variables resulting in choosing the output of the confidential program for any input among all executed programs. We fulfill (ii), since there is essentially just one obfuscated program (comprising of all programs), ie. any two programs $P'$ and $P$ map to the same obfuscated program.

Running entire programs and selecting the result is not ideal. When choosing $k$ programs with similar running times or computational costs, the running time increases by a factor of $k$. But an attacker only has to choose one out of $k$ programs, i.e., the attacker breaks the ``encryption'' with probability $1/k$ by guessing.  We would like to have at least an exponential relationship in the computational cost and the difficulty to guess, e.g., $1/2^k$. The key idea to reach this goal is to apply selector bits per statement rather than picking entire programs.

\subsection{Statement Level} \label{sec:sta}
An attacker should not be able to infer a specific statement, even given that he knows all prior and all following statements. Our strategy is analogous as for (entire) programs: In order to disguise the statement $s$ we create a set of $k-1$ `misleading' statements  $M_i$. Then all statements $\{M_0,...,M_{k-1},s\}$ are randomly permuted. After that they are executed in that order. We choose the correct result, ie. that of the confidential variable $s$, using binary selector variables $b_i$. More precisely, we compute the \emph{combining statement}, which is defined as follows:
\bde\label{def:comb}
The combining statement for a set of statements $\mathbb{S}=\{S_C,M_0,...,M_{k-1}\}$ with confidential statement $S_C$ and misleading statements $M_i$ is given by 
 $$ENC(r^*):= \sum_{s \in \mathbb{S}} ENC(b_s)\cdot ENC(r_s)$$ where $r_s$ is the result of statement $s$ and $b_s$ are binary (selector) variables, such that $b_{s}=1$ for $s=S_C$ and 0 otherwise.
\ede 
 
For example, for the statement $c:=MUL(a,a)$, we could choose a misleading statement $ADD(a,a)$ and use the combining statement $c':=MUL(a,a) \cdot b_0 + ADD(a,a) \cdot b_1$ with binary selector variable $b_0=1-b_1$ being 1 and $b_1$ being 0. Note, that since $b_0=1$ and $b_1=0$ we have $c=c'$. If the selector variables $b_0,b_1$ as well as other variables and subexpressions are encrypted, i.e., we compute $ENC(c'):=ENC(MUL(a,a))\cdot ENC(b_0) + ENC(ADD(a,a))\cdot ENC(b_1)$ (with $\cdot,+$ operating on encrypted data), then an attacker is not able to discover the true statement without knowing $b_0$ (or $b_1$).\\
In contrast to program level security (Section \ref{sec:pro}), we just execute a single but larger program. Generally, for the same level of security, this single program is much more compact (and efficient to evaluate) than the concatenation of all programs used for program security. Roughly speaking, for programs of length $n$, where each statement could be assigned one out of $k$ expressions, the number of feasible programs is $k^n$. The concatenation of all misleading statements is of length $k\cdot n$, since for each of the $n$ statements there are $k$ options. This concatenation covers all $k^n$ feasible programs. Therefore, by executing $k\cdot n$ rather than $n$ statements, an attacker has to choose from $k^n$ options. This polynomial relationship between computational costs and number of possible programs is much better than the linear relationship presented in the prior Section \ref{sec:pro}. Therefore, we focus on statement level security. 


So far, we have focused on obfuscating a single confidential statement by combining its result and the result of several misleading statements. In principle, one could also generate ``misleading combining statements'' consisting of misleading statements only, ie. the entire statement is irrelevant. Thus, these combining statements merely serve the purpose of confusing the attacker (and hiding the true length of the confidential program). In fact, such `misleading combining statements'' might make comprehension more difficult. But they do not make the life of the attacker harder, when it comes to obtaining a program with the same input-output behavior. The attacker does not have to guess any misleading statement correctly, since all of them have no impact on the output. 

Still, the number of possible expressions $k$ can be exceedingly large, rendering the execution of $k\cdot n$ statements practically infeasible.
Therefore, we can only choose a subset of all possible statements as discussed next.




\subsection{Program Classes} \label{sec:procla}
There is an interesting relationship between an obfuscated program $P_O$ and program classes $\mathbb{C}_{C}$ containing the confidential program $P_C$. An obfuscated program $P_O$ using binary selector variables defines a program class $\mathbb{C}_{C}$, i.e., a set of programs, containing the confidential program that could have been mapped by the obfuscator to create $P_O$. 

For example, the confidential program $P_C=\{r:=MUL(a,a)\}$ could have been mapped to the encrypted program $P_O:=\{r0:=MUL(ENC(a),ENC(a))); r1:=ADD(ENC(a),ENC(a));r2:= r0\cdot ENC(b_0)+r1\cdot ENC(b_1) \}$, where $MUL,ADD$ compute the multiplication and addition of plaintexts using encrypted values. The encrypted program $P_)$ defines the program class $\mathbb{C}_{C}$ consisting of $\{MUL(a,a), ADD(a,a)\}$. The program class defined by the encrypted program depends on the details of the obfuscation algorithm, its parameters and the input program. Intuitively, large program class $\mathbb{C}_{C}$ seem preferable, since they enlarge the search space for the attacker. Indistinguishability obfuscation ensures that an attacker seeing the obfuscated program cannot determine any information about which program of all programs in the program class $\mathbb{C}_{C}$ is the confidential program. But indistinguishability obfuscation alone does not necessarily give any practical security guarantees, if the program class $\mathbb{C}_{C}$ is known to the attacker and there exist many programs that are close to the confidential program. For example, if all programs in the program class $\mathbb{C}_{C}$  (including the confidential program) yield (almost) the same output on any input then an attacker can pick any program from the class to get (almost) the same functionality. 
Moreover, it might be possible to exclude many programs from the program class $\mathbb{C}_{C}$ through simple reasoning that might not even require background knowledge. Certain programs might be executable and yield valid outputs, but their semantics might seem unreasonable. For example, a program that returns the same output irrespective of the input. An attacker can safely ignore such programs therefore reducing the size of the program class containing the confidential algorithm. It is generally not clear how to best choose the program class so that the attacker cannot easily eliminate many programs from the class with more or less simple analysis or background knowledge. There are many factors that come into play. We discuss mechanisms to create obfuscated programs in the next section.


\section{Obfuscation Metrics} \label{def:me}
Our metrics primarily follow (but enhance) the metrics for (standard) obfuscation~\cite{coll98}. We introduce a new metric that characterizes the quality of a program class.\\
\noindent\textbf{Code potency}: Obfuscated code can be assessed using traditional code complexity measures, e.g., based on control flow and data access. One might also use parameters from the obfuscation process itself, which illustrate the complexity of the encrypted code. For example, if we choose (up to) $k$ misleading statements for each statement, we might use a metric such as a ``mislead factor''. This is analogous to the security parameter for encryption.\\
\noindent \textbf{Resilience}: This is the ability to withstand attacks using automated tools (maybe given some background knowledge). For example, an attacker could try to reconstruct the program using known-plaintext attacks, i.e., he knows the input of the function and a few outputs. The attacker could then simply try each program from the encryption space and return all matching ones. Another example involves knowledge of coding guidelines used in the program or having unencrypted code of the same programmer at hand. Programmers typically have their own style that might be comparable to a signature identifying the programmer. Thus, an attacker could search for patterns in the encrypted code that match the coding guidelines (or the patterns matching the programmer's coding style). This could help to reduce the search space of possible non-obfuscated programs given an obfuscated program. Two metrics have been proposed in the literature~\cite{coll98}: i) programmer effort to construct an automatic de-obfuscator; ii) de-obfuscator effort: execution time and space required by the de-obfuscator.\\
One might also consider a metric for measuring \emph{potency reduction}, stating how effective a de-obfuscator is. For example, for statement level security it could measure the ``mislead factor reduction'', i.e., the (average) number of misleading (or junk) statements that the de-obfuscator could identify per statement that has been obfuscated.\\
\noindent \textbf{Program class quality}: On a high level a program class should contain programs that are sufficiently different from the confidential program, such that knowing an obfuscated programs is of no value to an attacker. Given a program from the program class, an attacker should not easily be able to judge whether or not it is the confidential program. Since our obfuscated program using encrypted selector variables discloses the program class, both aspects are important. Giving a formal definition of a program class is difficult, since it depends on the semantics of a program. However, we might reason about the attackers capabilities and define the quality of the program class in terms of the probability estimates $p(P)$ of a rational attacker that a specific program is confidential. A rational attacker would leverage all its knowledge about the confidential program, the obfuscation algorithm etc. to compute a probability distribution $p(P)$ stating the probability estimate that a program $P$ is confidential. To determine the quality of the program class given the distribution of an attacker we define the rank $r$ of a program $P$ as the number of programs $P'$ with $p(P')\leq p(P_C)$. 
We define the quality of program class $\mathbb{C}_{C}$ with a subset $\mathbb{S} \subseteq  \mathbb{C}_{C}$ of programs such that all of the programs in $\mathbb{S}$ are considered confidential as $Q(\mathbb{C}_{C}):=1-1/r$ with $r$ being the minimum rank of any program $P \in \mathbb{S}$. The quality ranges between 0 and 1. It is 0 if the attacker correctly assigns a (or the) confidential program the largest probability to be confidential. It is 1 if the attacker believes it is the least likely. The motivation is that a very reasonable behavior for an attacker given the probability distribution $p(P)$ is to perform an in depth investigation of the programs sorted by their probability estimate. For example, assuming the attacker has access to some plaintexts (input and output), he might check one program after the other, whether its input-output behavior matches the known plaintexts. Thus, the time needed to infer the confidential program depends directly on our quality measure. Defining the set of confidential programs $\mathbb{S}$ is application dependent. For example, given one confidential program $P_C$ one might define the set $\mathbb{S}$ as all programs having the same input-output behavior for all inputs, ie. $\mathbb{S}=\{P|\forall x:  P(x)=P_C(x)\}$.\\
\noindent \textbf{Stealth} refers to difficulty to de-obfuscate programs by humans, meaning that junk statements should not be easily discoverable by a human. The ability to discover them significantly depends on the knowledge a human has about the program (and surrounding statements).\\
\noindent \textbf{Execution cost}: For instance, the overhead factor being the ratio of the (average) execution time of non-obfuscated and the obfuscated program.

\section{Practical Obfuscation on the Statement Level } \label{sec:praob}
Executing all possible (simple) expressions for a real-world programming language instead of just a single one for each statement in a program is infeasible for even moderately large programs. Thus, we introduce a weaker form of security by adding fewer misleading statements in order to improve performance. Choosing these misleading statements in a way that is hard to disguise is non-trivial.  One might have to take into account existing patterns in code. and, potentially, remove these coding patterns to create a more uniformly looking encrypted code that is less sensitive to statistical attacks (Section \ref{sec:uni}). We aim at schemes that make it possible to strike a balance between several metrics, in particular security and performance. We discuss metrics for obfuscation and security (Section \ref{def:me}) as well as what kind of information of a real-world program might be worth to hide (Section \ref{sec:goals}).


\subsection{Preprocessing - Source Code Uniformization} \label{sec:uni}
Before obfuscating source code, we might modify the code through several preprocessing operations that do not change the semantics of the code but make it more `uniform' in the sense that patterns consisting of potentially multiple statements are removed. We might also  avoid generating highly unlikely patterns. Source code might contain patterns due to preferences of a programmer or due to the nature of the problem. Programmer preferences might be found in certain command constructs used to solve frequently reoccurring tasks. Patterns can be mined and used for statistical attacks. Assume the obfuscated statement contains two types of comparison expressions $a\neq b$ and $a=b$ out of which one is misleading. Assume we know that a programmer frequently almost never uses $`\neq'$ but almost uses `not' $`\not'$ and $`='$ to express inequality, et. $\not (a\neq b)$. This information makes it easier to find non-obfuscated code. In another example, a programmer might also prefer to state a loop condition involving integer $I$ and loop variable $x$ as $x\leq I-1$ rather than $x<I$. The former condition $x\leq I-1$ is expressed using two statements, $J:=I-1$ and $x\leq J$. Therefore, when trying to determine a statement out of a set of misleading statements, and seeing a condition $x\leq J$, an attacker might investigate the preceding statements to check whether one of them contains a statement of the form $J:=I-1$, if so, this is an indication that the code contains $x\leq I-1$.  Both cases are easy to detect automatically in source code of typed languages. We can enforce using, eg. '$\cdot 2$' and $x<I$  by rewriting the code without changing the semantics. In fact, some cases might even be covered by compiler optimizations. This might reduce the possibilities for an attacker to exploit prior knowledge about certain programming preferences.

\subsection{Misleading Code Generation} \label{sec:mis}
An essential question for obfuscation relates to the choice of misleading statements to satisfy the metrics in Section~\ref{def:me}. Several considerations apply:\\
\noindent \textbf{Number of misleading statements}: Choosing relatively few (misleading) statements might cause only little overhead due to obfuscation but put security at jeopardy.\\
\noindent \textbf{Execution costs of a misleading statement}: It may be preferable to choose misleading statements that do not require a lot of computation. For example, rather than executing a `sort' command on a list as misleading statement, one might remove the first element. The first command requires $O(l \log l)$ time for a list of length $l$, whereas the amortized time for the second command is only $O(\log l)$~\cite{sch15ab}.\\
\noindent \textbf{Similarity of misleading and confidential statements}: One strategy is to choose misleading statements independently of existing source code. However, independently chosen statements might be easy to identify, since they might appear as outliers. For example, given a complex function of many simpler mathematical functions, such as sine, square root, division and so forth, it is not recommended to add misleading statements manipulating strings. It seems likely that a string expression is identified as not being part of the confidential program. To make the junk statements hard to identify, potentially given some knowledge about the application or source code structure, it seems favorable to choose misleading code that is similar to the actual source code.\\
\noindent \textbf{Intermediate code vs. source code}: One might add statements for intermediate code (e.g., Java bytecode) or plain source code. Generally, in our scenario only intermediate code is given to an untrusted party. An attacker typically decompiles it. Modifying intermediate code might cause decompilers to fail. Therefore, an attacker might have to work on intermediate code itself, which is considerably harder. Compilers can also generate intermediate code according to different metrics, for example with the goal to minimize the amount of code. In our case, in order to minimize the absolute performance impact of obfuscation, it seems preferable to generate short programs. As an example, if we add three junk statements per statement in a short program, the absolute increase in running time is less than if we do the same for a long program. Furthermore, intermediate code might remove certain coding patterns that could help in statistical attacks (see Section \ref{sec:uni}). Thus, if our obfuscation technique does not account for dependencies, an attacker might be able to rule out some misleading statements. In intermediate code, such patterns might be removed.\\
\noindent \textbf{Simplifying complex expressions}: One has to decide which constructs in the source code should be enhanced by misleading `statements'. Depending on the programming language, statements are allowed to vary a lot in complexity: they could be decomposable into many simple expressions. Disguising a complex statement as a whole, also requires a statement of similar complexity (otherwise an attacker might assume that very complex statements are non-junk). Trying to create complex misleading statements seems rather challenging, since a single subexpression that appears as outlier might be sufficient for an attacker to disguise the entire complex statement as `misleading'. Furthermore, in the most extreme case an entire program is expressed in one statement, which results in similar disadvantages as discussed when obfuscating entire programs compared to individual statements (Section \ref{sec:pro}). Therefore, it might be preferable to create a uniform source code representation of simple statements, such as three-address codes and obfuscate them.\\
\noindent \textbf{Syntactic and semantic correctness}: For three-address code, a junk statement comprises of an operation and two operands. Clearly, the operation must be executable for the operands. This implies that the operands must have a type that is supported by the operation. But the values of the operands should also be supported. For example, when adding a division as a junk statement, the divisor should not be zero.\\
\noindent \textbf{Operation and operands obfuscation}: We can create misleading statements by using the same operations but different operands or different operations with the same operands or we can vary both, e.g., for the statement $a\cdot b$, we could add  $a+b$ or $a-b$ to disguise the operation only and $x\cdot y$ or $e\cdot f$ to disguise the operands only. In order to combine both, we might create temporary variables involving several variables with selector variables and then we could use these temporary variables in various operations. This yields the largest search space for an attacker. For example, we can set temporary variable $t1$ to be  $t1:=b1\cdot a+(1-b1)\cdot x$ and   $t2:=b2\cdot b+(1-b2)\cdot y$ for selector variables $b1,b2$. We can use these temporary variables in statements $t1\cdot t2$ and $t1+t2$. This yields a total of eight possibilities of statements. More generally, if both temporary variables can be chosen out of $k$ variables and the operations can be chosen out of $o$ operations, we get $k^2\cdot o$ possibilities.\\
\noindent \textbf{Patterns and statement dependency}: Statements are generally not independent. Given a single statement the probability of the next statement is not uniform across all statements. Parts of the dependency are due to the data type, that is, only certain statements are feasible for certain types. For example, given we know a statement performs string concatenation, it seems more likely that among the next statements there is another statement for string operation rather than a trigonometric function. There might be common patterns consisting of a sequence of a few statements that are more likely than others in general or depending on a specific application domain or programmer. Finding the optimal strategy of the attacker and the obfuscator to use (or counteract) the knowledge of pattern could be cast as the problem of finding a mixed-strategy Nash equilibria, eg. we look at the obfuscator and the attacker as players. A formal treatment of Nash equilibria is beyond the scope of this work and we refer to text books on game theory covering these concepts, eg. \cite{cam10}. To get some intuition, consider the following scenario. There are $n$ different statements $\mathbb{S}$ that are not uniformly distributed. There is one frequent statement $F \in \mathbb{S}$ with (large) likelihood $p_l$ and every other of the $n-1$ statements has the same probability $1/(n-1)$, ie. we have $p_l\gg 1/n$. Assume the strategy of the obfuscator is to generate one misleading statement for each statement $s_C$ belonging to the confidential code $C$ by choosing it uniformly at random from $\mathbb{S}\setminus s_C$, ie. it does not choose the misleading statement to be the same as the confidential statement, since this provides no protection. Say an attacker wants to choose the correct statement for a pair of two statements $s_C,M$ with $M$ being the misleading statement and $s_C$ a statement of the confidential program. He would behave as follows to maximize the expected number of correctly inferred statements:  If one of the two statements $s_C,M$ equals $F$, he guesses $F$. If both statements do not equal $F$ he chooses randomly among the two. Let us compute the probability that an attacker correctly infers a statement given two statements $s_C,M$ of a large program. The attacker is always correct if the true statement $s_C$ is $F$, which happens for a fraction $p_l$ of all statements. If the true statement is not $F$ but the misleading statement is $F$ the attacker will always predict wrongly. If the misleading statement is not $F$ and the true statement is not $F$, it predicts correctly with probability $1/2$. Thus, the expected number of correctly predicted statements is $p_l+(1-p_l)\cdot (1-1/n)\cdot 1/2$. Now, assume that our obfuscator is aware of the true distribution of statements as well. Assume that it always chooses $F$ as misleading statement if $s_C\neq F$. If the attacker does not alter its strategy, the expected number of correctly guessed statements is now only $p_l$. However, we could do better using a different strategy.

\section{Empirical Evaluation}
In Section \ref{sec:mis} we introduced several general aspects for misleading code generation. In this section, we discuss some points in more detail and further provide a short evaluation by programmers. A key aspect is the structure of the code: How much dependency is there across statements? Does code contain patterns that must be taken into account when generating code? The second question is discussed next.

\subsection{Patterns in Source Code} \label{sec:pat}
Patterns in source code might facilitate the decoding of obfuscated code using encryption as discussed in the beginning of Section \ref{sec:praob}. There is a sheer endless number of patterns one could look for, ranging from high-level design patterns to the use of individual operators. Since we operate primarily on a statement level, we compared the frequencies of patterns on a lower level. We looked at the distribution of operators, i.e., we counted how often an operator appeared in the code, since simple statements essentially consist of an operator and one or two variables.  An even more particular statement involves an integer constant and an operator. We conjectured that the usage of integer constants might vary significantly across programmers, for example the usage of $\leq n-1$ vs. $<n$. Therefore, we looked at how integer constants are used within binary expressions, specifically what values they have and with what operator they are used with. In order to get some intuition about more complex patterns, we looked at the composition of binary and unary expressions: What are the type of expression(s) and the operation that constitute a binary (or unary) expression? 

For comparison, we chose five data mining frameworks implemented in JAVA namely WEKA, ELKI, JSAT, Java-ML and Spmf\footnote{See \url{http://www.cs.waikato.ac.nz/ml/weka/}; \url{http://elki.dbs.ifi.lmu.de/}; \url{https://github.com/EdwardRaff/JSAT} and for GUI \url{.../JSATFX}; \url{see http://java-ml.sourceforge.net/}; \url{http://www.philippe-fournier-viger.com/spmf/}}. We compared all clustering algorithms per framework together (see Table~\ref{tab1}) as well as particular algorithms, namely $k$-Means  and OPTICS clustering (for k-Means see Tables~\ref{tab3}), to get a more robust estimate. We also compared the GUI parts of the frameworks (see Table \ref{tab2}) to detect variation across different applications.  \\


The less uniform code is across programmers and application domains, the more important it is to take particular coding patterns into account. When looking at the standard deviation and the mean for an application domain on its own (Tables \ref{tab2} and \ref{tab3}) for all the structural patterns, we see that the mean is considerably larger than the standard deviation.  This shows limited dispersion and a certain uniformity. Therefore, knowing that code stems from a particular framework helps in predicting a statement but only to a rather limited degree. Generating misleading statements according to such a distribution seems helpful, i.e., we might choose misleading statements such that the overall distribution roughly stays the same. When comparing the means for the same patterns of different application domains, we see stronger variation. For example, the mean for the increment `posIncrement NameE' for GUI and clustering varies by a factor of two. Still, the differences are not that large that a decoding of 
statements would be easily possible by merely knowing that we consider a specific type of application. 
Judging from the usage of integer constants and operators, there are strong deviations for a few cases. For example, the plus operator occurs almost three times more often in WEKA than in JSAT or JavaML. JSAT does not use integer constants except divisions by 2, ELKI in turn never uses divisions by 2. This hints at programmer preferences. Given the strong variation for some examples, it seems advisable to create misleading commands based on patterns observed in the code.

\begin{table}[htp!]
	\begin{center} \resizebox{0.5\textwidth}{!}{\begin{tabular}{|l |l |l |l |l |l ||l|l|}\hline
				& \multicolumn{ 5 }{|c||}{Frameworks} & Mean & Std  \\ \hline
				Pattern &Weka&Elki&JSAT&Spmf&JavaML&&\\ \hline
				\hline\multicolumn{ 8 }{|c|}{Structural Pattern: Binary and Unary Expression } \\ \hline
				posIncrement NameE & 9\tiny{ (446)} & 7\tiny{ (316)} & 13\tiny{ (369)} & 10\tiny{ (37)} & 17\tiny{ (285)}&11.1&3.4\\ \hline
				\footnotesize{BinaryE plus StringLiteralE} & 8\tiny{ (387)} & 3\tiny{ (152)} & 0\tiny{ (13)} & 4\tiny{ (15)} & 0\tiny{ (7)}&3.2&2.8\\ \hline
				assign NameE & 7\tiny{ (342)} & 10\tiny{ (441)} & 10\tiny{ (285)} & 9\tiny{ (32)} & 10\tiny{ (165)}&8.9&1.1\\ \hline
				assign MethodCallE & 7\tiny{ (339)} & 9\tiny{ (410)} & 7\tiny{ (191)} & 8\tiny{ (31)} & 6\tiny{ (94)}&7.2&1.2\\ \hline
				NameE less NameE & 4\tiny{ (195)} & 5\tiny{ (251)} & 6\tiny{ (172)} & 2\tiny{ (7)} & 4\tiny{ (62)}&4.1&1.5\\ \hline
				\footnotesize{assign ObjectCreationE} & 3\tiny{ (171)} & 1\tiny{ (44)} & 1\tiny{ (34)} & 3\tiny{ (13)} & 2\tiny{ (35)}&2.2&1.1\\ \hline
				\hline\multicolumn{ 8 }{|c|}{Integer Constants in Binary Expression } \\ \hline
				plus 1 & 22\tiny{ (114)} & 19\tiny{ (78)} & 22\tiny{ (61)} & 3\tiny{ (1)} & 17\tiny{ (31)}&16.6&6.8\\ \hline
				minus 1 & 17\tiny{ (90)} & 23\tiny{ (95)} & 21\tiny{ (57)} & 31\tiny{ (9)} & 40\tiny{ (72)}&26.3&8.4\\ \hline
				greater 0 & 16\tiny{ (86)} & 11\tiny{ (45)} & 9\tiny{ (25)} & 7\tiny{ (2)} & 9\tiny{ (16)}&10.3&3.2\\ \hline
				equals 0 & 12\tiny{ (63)} & 12\tiny{ (50)} & 5\tiny{ (13)} & 17\tiny{ (5)} & 5\tiny{ (9)}&10.1&4.8\\ \hline
				notEquals 0 & 10\tiny{ (50)} & 1\tiny{ (3)} & 0\tiny{ (0)} & 7\tiny{ (2)} & 2\tiny{ (4)}&3.8&3.7\\ \hline
				divide 2 & 4\tiny{ (21)} & 0\tiny{ (0)} & 7\tiny{ (19)} & 0\tiny{ (0)} & 11\tiny{ (20)}&4.4&4.3\\ \hline
				\hline\multicolumn{ 8 }{|c|}{Operators} \\ \hline
				assign & 32\tiny{ (1360)} & 28\tiny{ (1106)} & 31\tiny{ (719)} & 31\tiny{ (100)} & 32\tiny{ (437)}&30.6&1.7\\ \hline
				plus & 29\tiny{ (1210)} & 19\tiny{ (748)} & 12\tiny{ (284)} & 19\tiny{ (61)} & 11\tiny{ (145)}&17.7&6.3\\ \hline
				less & 11\tiny{ (479)} & 12\tiny{ (478)} & 20\tiny{ (458)} & 11\tiny{ (36)} & 21\tiny{ (289)}&14.9&4.4\\ \hline
				minus & 6\tiny{ (242)} & 6\tiny{ (260)} & 6\tiny{ (142)} & 7\tiny{ (24)} & 10\tiny{ (130)}&7.0&1.4\\ \hline
				equals & 5\tiny{ (228)} & 6\tiny{ (242)} & 7\tiny{ (155)} & 12\tiny{ (39)} & 3\tiny{ (45)}&6.6&2.9\\ \hline
				greater & 4\tiny{ (166)} & 4\tiny{ (178)} & 3\tiny{ (72)} & 2\tiny{ (7)} & 4\tiny{ (54)}&3.5&0.8\\ \hline
			\end{tabular}} \caption{Sorted patterns for all clustering algorithms with relative(\%) and absolute frequencies in brackets} \label{tab1} \end{center} 
	\end{table}
	
	\begin{table}[htp!]
		\begin{center} \resizebox{0.5\textwidth}{!}{\begin{tabular}{|l |l |l |l |l ||l|l|}\hline
					& \multicolumn{ 4 }{|c||}{Frameworks} & Mean & Std  \\ \hline
					Pattern &Weka&Elki&JSAT&Spmf&&\\ \hline
					\hline\multicolumn{ 7 }{|c|}{Structural Pattern: Binary and Unary Expression } \\ \hline
					assign MethodCallE & 7\tiny{ (1871)} & 5\tiny{ (31)} & 2\tiny{ (2)} & 6\tiny{ (39)}&5.0&2.0\\ \hline
					\footnotesize{assign ObjectCreationE} & 6\tiny{ (1501)} & 11\tiny{ (69)} & 8\tiny{ (9)} & 11\tiny{ (73)}&9.0&2.2\\ \hline
					assign NameE & 6\tiny{ (1423)} & 7\tiny{ (43)} & 5\tiny{ (6)} & 6\tiny{ (37)}&5.8&0.5\\ \hline
					\footnotesize{NameE notEquals NullLiteralE} & 5\tiny{ (1183)} & 7\tiny{ (47)} & 3\tiny{ (3)} & 1\tiny{ (9)}&4.0&2.2\\ \hline
					\footnotesize{BinaryE plus StringLiteralE} & 4\tiny{ (1084)} & 0\tiny{ (0)} & 0\tiny{ (0)} & 3\tiny{ (19)}&1.8&1.8\\ \hline
					posIncrement NameE & 4\tiny{ (1048)} & 1\tiny{ (4)} & 13\tiny{ (14)} & 4\tiny{ (23)}&5.2&4.5\\ \hline
					\hline\multicolumn{ 7 }{|c|}{Integer Constants in Binary Expression } \\ \hline
					greater 0 & 19\tiny{ (570)} & 40\tiny{ (25)} & 0\tiny{ (0)} & 1\tiny{ (1)}&15.1&16.3\\ \hline
					minus 1 & 13\tiny{ (401)} & 6\tiny{ (4)} & 0\tiny{ (0)} & 13\tiny{ (9)}&8.1&5.4\\ \hline
					plus 1 & 11\tiny{ (344)} & 8\tiny{ (5)} & 0\tiny{ (0)} & 1\tiny{ (1)}&5.2&4.7\\ \hline
					equals 0 & 11\tiny{ (340)} & 11\tiny{ (7)} & 0\tiny{ (0)} & 12\tiny{ (8)}&8.5&4.9\\ \hline
					divide 2 & 9\tiny{ (267)} & 0\tiny{ (0)} & 61\tiny{ (11)} & 0\tiny{ (0)}&17.5&25.5\\ \hline
					greaterEquals 0 & 5\tiny{ (163)} & 5\tiny{ (3)} & 0\tiny{ (0)} & 1\tiny{ (1)}&2.9&2.2\\ \hline
					\hline\multicolumn{ 7 }{|c|}{Operators} \\ \hline
					assign & 37\tiny{ (8687)} & 40\tiny{ (240)} & 18\tiny{ (17)} & 28\tiny{ (167)}&30.8&8.8\\ \hline
					plus & 22\tiny{ (5032)} & 11\tiny{ (63)} & 17\tiny{ (16)} & 14\tiny{ (85)}&15.8&4.0\\ \hline
					equals & 8\tiny{ (1884)} & 12\tiny{ (73)} & 1\tiny{ (1)} & 15\tiny{ (89)}&9.0&5.2\\ \hline
					notEquals & 7\tiny{ (1736)} & 11\tiny{ (66)} & 3\tiny{ (3)} & 5\tiny{ (29)}&6.6&3.0\\ \hline
					less & 5\tiny{ (1228)} & 2\tiny{ (14)} & 17\tiny{ (16)} & 3\tiny{ (20)}&7.0&5.9\\ \hline
					minus & 5\tiny{ (1140)} & 3\tiny{ (15)} & 14\tiny{ (13)} & 2\tiny{ (15)}&5.9&4.7\\ \hline
				\end{tabular}} \caption{Sorted patterns for GUI with relative(\%) and absolute frequencies in brackets} \label{tab2} \end{center} 
		\end{table}
		
		\begin{table}[htp!]
			\begin{center} \resizebox{0.5\textwidth}{!}{\begin{tabular}{|l |l |l |l |l |l ||l|l|}\hline
						& \multicolumn{ 5 }{|c||}{Frameworks} & Mean & Std  \\ \hline
						Pattern &Weka&Elki&JSAT&Spmf&JavaML&&\\ \hline
						\hline\multicolumn{ 8 }{|c|}{Structural Pattern: Binary and Unary Expression } \\ \hline
						assign MethodCallE & 12\tiny{ (65)} & 10\tiny{ (7)} & 7\tiny{ (6)} & 11\tiny{ (7)} & 0\tiny{ (0)}&8.1&4.3\\ \hline
						posIncrement NameE & 8\tiny{ (46)} & 13\tiny{ (9)} & 8\tiny{ (7)} & 8\tiny{ (5)} & 20\tiny{ (13)}&11.6&4.7\\ \hline
						assign NameE & 6\tiny{ (33)} & 13\tiny{ (9)} & 16\tiny{ (13)} & 11\tiny{ (7)} & 14\tiny{ (9)}&12.0&3.3\\ \hline
						\footnotesize{BinaryE plus StringLiteralE} & 5\tiny{ (27)} & 0\tiny{ (0)} & 0\tiny{ (0)} & 3\tiny{ (2)} & 0\tiny{ (0)}&1.6&2.1\\ \hline
						\footnotesize{less MethodCallE NameE} & 4\tiny{ (23)} & 3\tiny{ (2)} & 0\tiny{ (0)} & 2\tiny{ (1)} & 5\tiny{ (3)}&2.6&1.7\\ \hline
						\footnotesize{plus IntegerLiteralE NameE} & 4\tiny{ (22)} & 0\tiny{ (0)} & 0\tiny{ (0)} & 0\tiny{ (0)} & 0\tiny{ (0)}&0.8&1.6\\ \hline
						\hline\multicolumn{ 8 }{|c|}{Integer Constants in Binary Expression } \\ \hline
						plus 1 & 40\tiny{ (29)} & 13\tiny{ (1)} & 25\tiny{ (2)} & 0\tiny{ (0)} & 0\tiny{ (0)}&15.4&15.3\\ \hline
						greater 0 & 18\tiny{ (13)} & 13\tiny{ (1)} & 0\tiny{ (0)} & 0\tiny{ (0)} & 33\tiny{ (1)}&12.7&12.4\\ \hline
						equals 0 & 7\tiny{ (5)} & 25\tiny{ (2)} & 0\tiny{ (0)} & 0\tiny{ (0)} & 67\tiny{ (2)}&19.7&25.2\\ \hline
						minus 1 & 7\tiny{ (5)} & 0\tiny{ (0)} & 25\tiny{ (2)} & 20\tiny{ (1)} & 0\tiny{ (0)}&10.4&10.3\\ \hline
						plus 2 & 5\tiny{ (4)} & 0\tiny{ (0)} & 0\tiny{ (0)} & 0\tiny{ (0)} & 0\tiny{ (0)}&1.1&2.2\\ \hline
						notEquals 0 & 4\tiny{ (3)} & 0\tiny{ (0)} & 0\tiny{ (0)} & 0\tiny{ (0)} & 0\tiny{ (0)}&0.8&1.6\\ \hline
						\hline\multicolumn{ 8 }{|c|}{Operators} \\ \hline
						assign & 36\tiny{ (175)} & 29\tiny{ (16)} & 44\tiny{ (32)} & 36\tiny{ (20)} & 44\tiny{ (21)}&37.9&5.7\\ \hline
						plus & 28\tiny{ (136)} & 15\tiny{ (8)} & 7\tiny{ (5)} & 15\tiny{ (8)} & 6\tiny{ (3)}&14.0&7.9\\ \hline
						less & 11\tiny{ (52)} & 20\tiny{ (11)} & 15\tiny{ (11)} & 13\tiny{ (7)} & 27\tiny{ (13)}&17.1&5.8\\ \hline
						minus & 7\tiny{ (35)} & 0\tiny{ (0)} & 14\tiny{ (10)} & 5\tiny{ (3)} & 4\tiny{ (2)}&6.1&4.5\\ \hline
						greater & 5\tiny{ (22)} & 2\tiny{ (1)} & 3\tiny{ (2)} & 4\tiny{ (2)} & 4\tiny{ (2)}&3.3&1.0\\ \hline
						equals & 4\tiny{ (19)} & 9\tiny{ (5)} & 4\tiny{ (3)} & 13\tiny{ (7)} & 4\tiny{ (2)}&6.8&3.5\\ \hline
					\end{tabular}} \caption{Sorted patterns for k-Means algorithm with relative(\%) and absolute frequencies in brackets} \label{tab3} \end{center} 
			\end{table}

\subsection{(De-)Obfuscation Case Study}
In this section we present examples of how simple obfuscated code might look like and we also perform a small-scale empirical evaluation with respect to the ability of humans to break the obfuscation, ie. the discover the confidential program. To the best of our knowledge, we are among the first to do such an empirical study providing some evidence on the value of obfuscation.
We (manually) obfuscated code for two simple tasks that are understandable by any novice programmer. The first task is to avoid division by zero by checking the value of a divisor and return some (error) value, if the divisor is zero:
\begin{verbatim} 
   if (y!= 0) then r:=x/y else r:=-9999
\end{verbatim}
\noindent The second task is to return the maximum of an array:
\begin{verbatim}
    m:=a[0]     			
    for(x=0;x<y;x=x+1)
      if m >= a[x] then m:=a[x]	
\end{verbatim}
We assumed that all constants are encrypted and appear as variables. The code of task one becomes: 
\begin{verbatim}
    if (y!= u) then r:=x/y else r:=v
\end{verbatim}

\noindent\underline{Obfuscation:}
There are many different options to obfuscate the code using encryption (see Section~\ref{def:me}). Discussing multiple techniques in detail is out of scope for this work. Therefore, we focus on a case study with simple manual code generation. We considered two levels of obfuscation ($L0$ and $L1$) with a different number of misleading statements and variables.\\

For the lower-level obfuscation, i.e., $L0$, we introduced two fake variables $w,z$. Thus, for obfuscation we allow any operand for the first task to be one of the variables $u,v,w,x,y,z$. For comparison operations, we allowed either a `non-equal' or a `smaller than' comparison. For mathematical operations, we used multiplication or division.  We did not introduce any additional fake statements.  Thus, for the $L0$ obfuscation, since variables occur at 5 places, this gives $5^5$ options for choosing variables and $2^2$ for choosing operations. Overall, this gives a total of 12,500 options. For the first simple assessment by humans, we limited the number of source code options, i.e., the program class $\mathbb{C}_{C}$, and rather presented complete code examples. More specifically, we chose 10 options for each task yielding 10 code examples chosen from $\mathbb{C}_{C}$. They were selected such that an ordinary software engineer could solve them without additional knowledge in a short amount of time. For example, we had an option containing a statement 'r:=z/z' 
that was easy to identify as most likely not being coded by a human. We figured that letting a programmer deal with encrypted code with binary selector variables is too complex. \\

\begin{figure}[!ht]
	\centerline{\includegraphics[width=\linewidth]{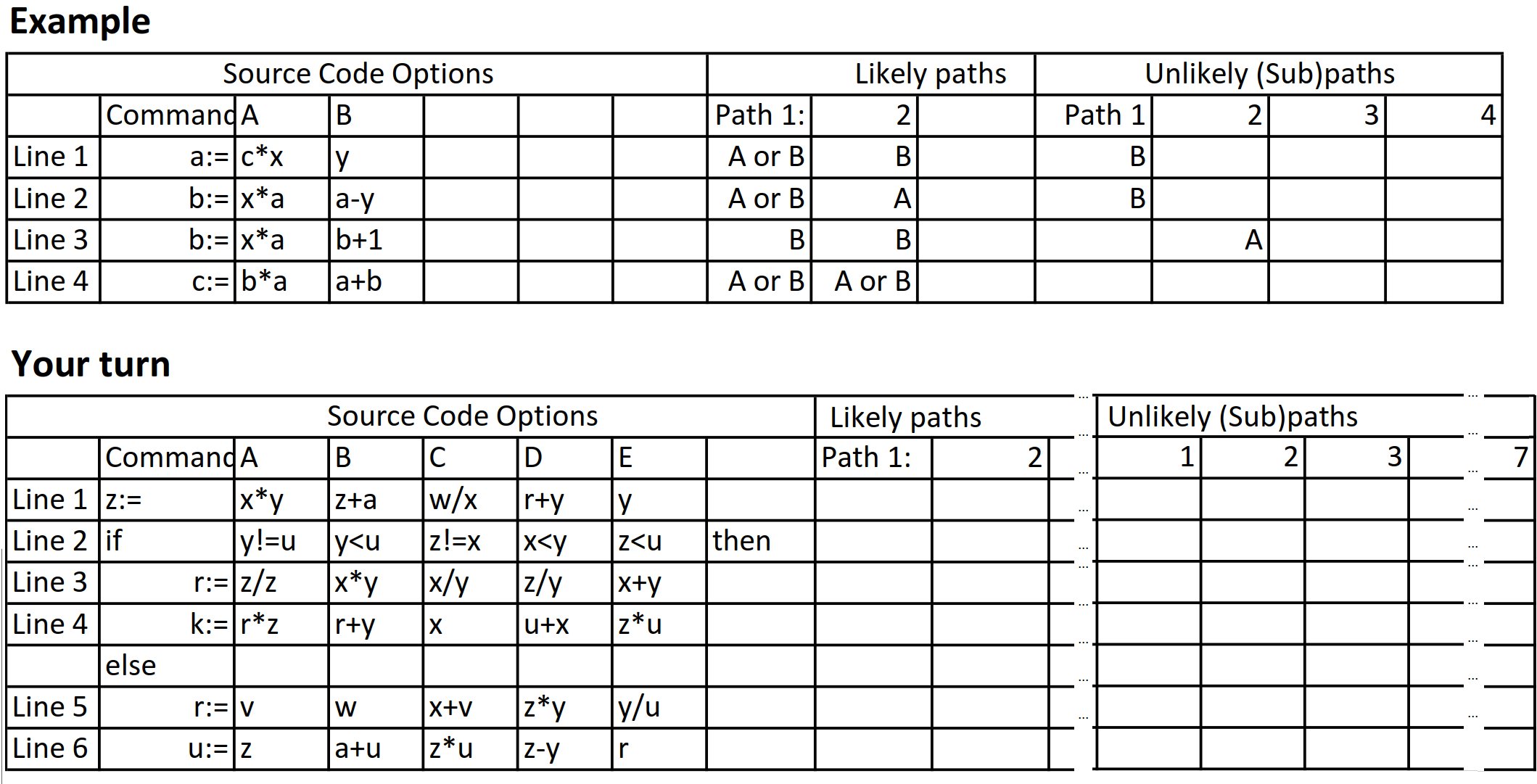}}
	\caption{Handouts to programmers for Simple Obfuscation L1}
	\label{fig:handoutL1}
\end{figure}

For the more thorough obfuscation ($L1$) we used three misleading combining statements and five simple statements per combining statement, ie. for each statement in the confidential code we added four more statemetns. We used 8 instead of 6 variables and additionally $+,-$ operators. In contrast to $L0$, where we only listed 10 options per task, we allowed for 5 options per 6 statements, yielding a total of $5^6 = 15,625$ options for the participants to choose from.  The handouts to programmers are listed in Figure \ref{fig:handoutL1}.
\\
\smallskip

\noindent\underline{Experimental Setup:}
We gave the programmers a short introduction ($\approx$ 15 minutes) about the obfuscation technique in general including examples. The programmers were not told what problem the source code solves, only that it is a common and simple task. We conducted two separate experiments one for $L0$ and for $L1$ obfuscation with different programmers. \\ 
Experiment 1: For $L0$ we used a total of 6 programmers. They were given a paper sheet with the source code options. They had 20 minutes to break the code. The participants had to rate how much sense an option makes to be the piece of code solving the problem. The answers had 5 options expressing different degree of confidence. The perfect answer would consist of one `Yes, that's it' and nine answers `No, it's not'. Since all options are syntactically correct, programmers needed some kind of understanding of the semantics of the program. Even given very good understanding, they might not be able to get the perfect answer since they lack background information about the task. Rating the source code options corresponds to the process of de-obfuscation, since programmers choose a probability distribution over all available options for confidential programs.\\ 
Experiment 2: For $L1$ we used 5 programmers (different from the first experiment). We only considered the first task dealing with division by zero. The handout to programmers is given in Figure \ref{fig:handoutL1}. In this case, it seems very difficult to identify the task. Therefore, we asked them to state unlikely and likely combinations of options for each statement. Put differently, rather than specifying for each statement just a single option they could rate several options as likely (or unlikely). Thus, they could state sets of options using multiple choices per statement. This corresponds to fixing paths (or traces) in the code (see the example shown in Figure \ref{fig:handoutL1}). They were also given 20 minutes to solve the task.

\noindent\underline{Results:}
Experiment 1: Overall programmers solved this task well, ie. they managed to de-obfuscate the code. Given the limited set of options this is not surprising. Four participants were indeed able to identify the correct option for the first task. The others ranked it as likely but together with at least one other option. On average, programmers were only able to reliably exclude two options. For the second task, there were two equally valid solutions.  One for computing the minimum of the array and one for the maximum. Four participants identified both options. One chose a wrong solution. One participant got one answer correct and rated several others equally likely, including the correct one. Programmers required 20 minutes or less for both tasks.\\

Experiment 2: Overall participants could not significantly narrow the search space. Often their intuition seemed not helpful, eg. the (statement) options rated as `likely' did not contain any statements belonging to the confidential code. We discuss identifying unlikely options first. All participants could correctly identify unlikely options. However, they were rather specific, ie. five or less pairs of unlikely options, eg. similar to the example in Figure \ref{fig:handoutL1}. Thus, none of the participants managed to reduce the search space by more than a factor of two. Identifying likely paths was even less successful: One participant did not mention any likely paths, one participant mentioned paths not sharing any common option with the solution. One participant mentioned paths with various options per statement but in fact only one of they had only one single option in common with the solution. One answer contained paths with two statements that were also part of the solution. Another answer contained paths, out of which only one had one statement in common with the solution. Thus, none of the participants was able to identify a set of paths that contained the solution.\\

In summary, the second task was significantly more difficult and the participants were not successful at de-obfuscating. Overall, we conclude that humans were not able to correctly and efficiently narrow down the search space to a reasonable number of program options. Although this might change given more time and experience. We also want to emphasize that we used only a toy example both with respect to the number of options generated and also with respect to code length. Thus, we believe that source code encryption is effective.

\section{Related Work}
There is a variety of approaches to avoid piracy, reverse engineering, and tampering \cite{nau13} such as license files, checksums  obfuscation, to name but a few. We only discuss obfuscation:\\
\noindent\underline{(De-)Obfuscation techniques:} There is a rich literature on source code obfuscation techniques, including general surveys (e.g., \cite{bal05}) as well as a more recent survey focusing on malware obfuscation techniques~\cite{you10} and a short discussion involving (indistinguishability) obfuscation \cite{beu16}. One common technique is to introduce opaque predicates to insert dead code. The idea is to have a control statement with two branches that always executes one of the branches. However, it should not be detectable by (static) analysis that the condition always evaluates to the same value. Clearly, dynamic analysis might yield some hints whether or not the code in one branch is dead, but there are theoretical limitations (as the problem is reducible to the halting problem) \cite{bea07}. Since we encrypt the condition and always execute both branches, it is impossible that code analysis provides any information about which branch is executed. In ordinary programs, adding fake or junk statements must be done with care because the program must remain correct. Thus, typically any change to the program by a junk statement that modifies the final outcome has to be undone, eg. by applying the inverse operation. However, little is known on how to actually create junk statements to disguise a human programmer. In the same work, obfuscation through dead code, void code, code duplication are briefly mentioned but with primary focus on avoiding detection by an automated de-obfuscater (rather than by a reverse engineer). An Intermediate Level Obfuscation Method \cite{dun14} gives some high level guidelines for obfuscation: The authors mention that constants can be calculated automatically because they might help to identify an algorithm. They also mention dead code generation focusing on alias analysis. Furthermore, they state briefly that dead code should be similar to the original executable code but without giving further insights. There is also work on adding junk to confuse a disassembler~\cite{lin03}. The junk statements should be unreachable and partial statements. Other than that, there is no discussion on what statements to choose. We are not limited to add junk statements and their inverses in our approach, since we can use selector variables to control whether or not a junk statement influences program state. Generally, the effectiveness of obfuscation techniques is still subject to study~\cite{anc07,cec14}. For example, one possible attack against obfuscation is frequency analysis using, e.g., pattern mining algorithms. Pattern mining has been employed for program comprehension \cite{tjo03} and reverse engineering \cite{maq04}. We provide an empirical analysis of source code patterns for similar problems across developers. These patterns could be used to create obfuscated statements that are harder to identify.

Recent work has identified programmers using abstract syntax trees with a surprisingly high success rate despite obfuscation of code \cite{cal15}. We do not attempt to identify programmers based on code, but we use code of the same programmer to support the reconstruction of obfuscated code. We use only small parts of the abstract syntax tree and compute also aggregate (i.e, pattern) statistics.

Obfuscation is also (ab)used by malware \cite{oka11} to circumvent intrusion detection systems by using polymorphic techniques. Though obfuscation is typically associated with source or machine code, obfuscation techniques have also been applied to data \cite{bak04}. Obfuscation has also been leverage to to construct an abstract state machine that enables computation on encrypted and non-encrypted data \cite{maz16}.

\noindent\underline{Metrics and goals:}
Metrics to assess code obfuscation have been discussed in the literature as well~\cite{coll98}, namely code potency (related to code complexity measures), resilience (against an automated de-obfuscator), stealth (easiness to spot obfuscated code by reverse engineer) and (additional) execution cost. We maintain the underlying ideas of these metrics and adjust them to our context. We also add a new metric `program class quality'.
In this work, we mainly focus on obfuscating functions. However, literature on obfuscation has discussed other directions as well, e.g., false refactoring to disguise class structure~\cite{bal05} or removing type information for Byte-code \cite{fok14}.

\noindent\underline{Circuit Privacy:} Traditional program obfuscation attempts disguise the computation (of a circuit) in one way or another. For instance, Indistinguishability obfuscation requires that given any two equivalent circuits $C0$ and $C1$ of similar size, the obfuscations of $C0$ and $C1$ should be computationally indistinguishable~\cite{gar13}. In functional encryption~\cite{gar13}, ciphertexts encrypt inputs $x$, and keys are issued for circuits $C$. Using a key to decrypt a ciphertext $ENC(C(x))$, yields the result of the circuit $C$ evaluated for $x$ but does not reveal anything else about $x$. Furthermore, no collusion of secret key holders should be able to learn anything more than the union of what they can each learn individually. Since the introduction of these concepts~\cite{gar13}, significant progress has been made (see \cite{hor15} for an overview). We argue that from a practical perspective, encryption at the source (or byte) code level is more meaningful than for Boolean circuits due to performance reasons. Given a processor 
supports $n$ commands each implemented using circuits of size $x \gg \log n$, it suffices to hide the command type rather than the circuit. 

One cannot hope to obfuscate arbitrary programs \cite{bar12} when requiring that a circuit should leak no information except its input and output behavior. Using a more relaxed notion for obfuscation, an obfuscated program may leak as much information as any other program with equivalent functionality~ \cite{gol07}. However, this makes information theoretic obfuscation impossible (in polynomial time).

Keeping the circuit private while changing any wire value in the circuit has also been investigated \cite{ish06}.
An attack can be detected and data can then be erased.

In his thesis Gentry~\cite{TGe09} gave a fully homomorphic encryption (FHE) scheme. He also discussed how to ensure circuit privacy by adding a large random error vector. 


\section{Conclusions}
Computing on encrypted data is close to being practical, e.g., using secure multi-party computation. In this paper, we have looked at the next step after protecting data privacy: Keeping algorithms confidential at the source code or bytecode level. We have shown that by adding misleading statements a high degree of protection can be achieved with a modest increase in computational complexity given that computation is carried out on encrypted data.
While our assessment showed the effectiveness of our approach, this is just one of the first steps and there are many interesting directions for future work in this domain.

\bibliographystyle{abbrv}
\bibliography{refs}

\end{document}